\documentclass[runningheads]{llncs}
\usepackage{graphicx}

\usepackage{tikz}
\usepackage{comment}
\usepackage{amsmath,amssymb} 
\usepackage{color}
\usepackage{multirow}
\usepackage{bbm}
\usepackage{array}
\usepackage{url}
\usepackage{booktabs}
\usepackage[accsupp]{axessibility}  

\begin{document}
\pagestyle{headings}
\mainmatter
\def\ECCVSubNumber{11}  

\title{CMC\_v2: Towards More Accurate COVID-19 Detection with Discriminative Video Priors} 

\titlerunning{CMC\_v2: COVID-19 Detection with Video Priors}
%
\author{Junlin Hou\inst{1} \and Jilan Xu\inst{1,3} \and Nan Zhang\inst{2} \and Yi Wang\inst{3} \and Yuejie Zhang\inst{1}* \and \\ Xiaobo Zhang\inst{4}* \and Rui Feng\inst{1,2,4}*}
\authorrunning{J. Hou et al.}
%

\institute{School of Computer Science, Shanghai Key Laboratory of Intelligent Information Processing, Fudan University, China\\
\and
Academy for Engineering and Technology, Fudan University, China\\
\email{\{jlhou18,jilanxu18,20210860062,yjzhang,fengrui\}@fudan.edu.cn}
\and Shanghai AI Laboratory, China\\
\email{wygamle@gmail.com}
\and Children’s Hospital of Fudan University, National Children’s Medical Center, Shanghai, China\\
\email{zhangxiaobo0307@163.com}\\
}
\maketitle

\begin{abstract}
This paper presents our solution for the 2nd COVID-19 Competition, occurring in the framework of the AIMIA Workshop at the European Conference on Computer Vision (ECCV 2022). In our approach, we employ the winning solution last year which uses a strong 3D Contrastive Mixup Classification network (CMC\_v1) as the baseline method, composed of contrastive representation learning and mixup classification. In this paper, we propose CMC\_v2 by introducing natural video priors to COVID-19 diagnosis. Specifically, we adapt a pre-trained (on video dataset) video transformer backbone to COVID-19 detection. Moreover, advanced training strategies, including hybrid mixup and cutmix, slice-level augmentation, and small resolution training are also utilized to boost the robustness and the generalization ability of the model. Among 14 participating teams, CMC\_v2 ranked 1st in the 2nd COVID-19 Competition with an average Macro F1 Score of 89.11\%.

\keywords{COVID-19 detection, Hybrid CNN-transformer, Contrastive learning, Hybrid mixup and cutmix}
\end{abstract}

\section{Introduction} 
\label{section:sec1}
The Coronavirus Disease 2019 SARS-CoV-2 (COVID-19), identified at the end of 2019, is a highly infectious disease, leading to an everlasting worldwide pandemic and collateral economic damage \cite{WHO}. Early detection of COVID-19 is crucial to the timely treatment of patients, and beneficial to slowdown or even break viral transmission. COVID-19 detection aims to identify COVID from non-COVID cases. Among several COVID-19 detection means, chest computed tomography (CT) has been recognized as a key component in the diagnostic procedure for COVID-19. In CT, we resort to typical radiological findings to confirm COVID-19, including ground glass opacities, opacities with rounded morphology, crazy-paving pattern, and consolidations \cite{chung2020ct}.
As a CT volume contains hundreds of slices, delivering a convincing diagnosis from these data demands a heavy workload on radiologists. Relying on manual analysis is barely scalable considering the surging increasing number of infection cases. Regarding this, there is an urgent need for accurate automated COVID-19 diagnosis approaches.

\begin{figure}[t]
\centering
\includegraphics[width=\textwidth]{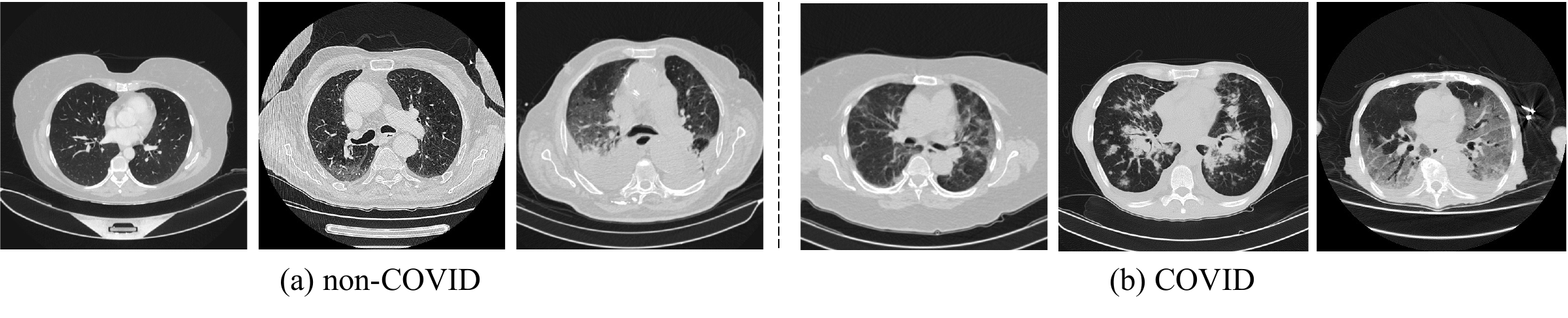}
\caption{Some examples of (a) non-COVID and (b) COVID cases from the COV19-CT-DB dataset. The non-COVID category includes no pneumonia and other pneumonia cases. The COVID category contains COVID-19 cases of different severity levels.}
\label{fig:intro}
\end{figure}

Recently, deep learning approaches have achieved promising performance in fighting against COVID-19. They have been widely applied to various medical practices, including the lung and infection region segmentation \cite{weakly,li2020artificial,Chen2020.02.25.20021568,tsinghua2020fourweek} as well as the clinical diagnosis and assessment \cite{wang2020automatically,wang2020a,song2020deep,HOU2021108005}. Though a line of works \cite{tsinghua2020fourweek,wang2020automatically,HOU2021108005} has been employed for COVID-19 detection via CT analysis and yielded effective results, it is still worth pushing its detection performance to a new level in a faster and more accurate manner for a better medical assistant experience. Improving this performance is non-trivial, since the inner variances between CT scans of COVID are huge and its differences with some non-COVID like pneumonia are easily overlooked. Specifically, CT scans vary greatly in imaging across different devices and hospitals (Fig. \ref{fig:intro}), and they share several similar visual manifestations with other types of pneumonia. Further, the scarcity of CT scans of COVID-19 due to regulations in the medical area makes these challenges even harder, as we cannot simply turn to a deep model to learn these mentioned characteristics with a big number of annotated scans from scratch.

To tackle these challenges, we exploit video priors along with the given limited number of CT scans to learn an effective feature space for COVID-19 detection, along with contrastive training and some hybrid data augmentation means for further data-efficient learning. Specifically, we employ the advanced 3D contrastive mixup classification network (CMC-COV19D, abbr. CMC\_v1) \cite{hou2021cmc}, the winner in the ICCV 2021 COVID-19 Diagnosis Competition of AI-enabled Medical Image Analysis Workshop \cite{kollias2021mia}, as a baseline. CMC\_v1 introduces contrastive representation learning to discover discriminative representations of COVID-19 cases. Besides, a joint training loss is devised by combining the classification loss, mixup loss, and contrastive loss. In this work, we propose CMC\_v2 by introducing the following mechanisms customized for 3D models. (1) To capture the long-range lesion span across the slices in the CT scans, we adopt a hybrid CNN-transformer model, i.e. Uniformer \cite{li2022uniformer} as the backbone network. The combination of convolution and self-attention reduces the network parameters and computational costs. It relieves the potential overfitting when deploying 3D models to small-scale medical datasets. Besides, we empirically show that initializing the model with 3D weights pre-trained on video datasets is promising as modeling the relationship among slices is critical for COVID-19 detection. (2) We develop a hybrid mixup and cutmix augmentation strategy to enhance the models' generalization ability. Due to the limited memory, a gather-and-dispatch mechanism is also customized for the modern Distributed DataParallel (DDP) scheme in Multi-GPU training. (3) We showcase both the 2D slice-level augmentation and the small resolution training bring improvements. By applying the intra-and-inter model ensemble \cite{hou2021cmc}, CMC\_v2 won the first prize in the 2nd COVID-19 detection challenge of the Workshop ``AI-enabled Medical Image Analysis – Digital Pathology \& Radiology/COVID19 (AIMIA)''. CMC\_v2 significantly outperforms the baseline model provided by the organizers by 16\% Macro F1 Score. 


The remainder of this paper is organized as follows. Section \ref{section:relatedwork} reviews related works. In Section \ref{section:method}, we first recap the CMC\_v1 network, the basis of CMC\_v2, and then introduce the newly proposed modules in CMC\_v2.
Section \ref{section:dataset} describes the COV19-CT-DB dataset used in this paper. Section \ref{section:experiments} provides the experimental settings and results. Section \ref{section:conclusion} concludes our work.

\section{Related Work}
\label{section:relatedwork}

\subsection{COVID-19 detection}
Numerous deep learning approaches have made great efforts to separate COVID patients from non-COVID subjects. Despite the binary classification, the task is challenging as the non-COVID cases include both common pneumonia subjects and non-pneumonia subjects.

The majority of deep learning approaches are based on Convolutional Neural Networks (CNN). \cite{wang2020a} was a pioneering work that designed a CNN model to classify COVID-19 and typical viral pneumonia. 
Song et al. \cite{song2020deep} proposed a deep learning-based CT diagnosis system (Deep Pneumonia) to detect patients with COVID-19 from patients with bacteria pneumonia and healthy people. 
Li et al. \cite{li2020artificial} developed a 3D COVNet based on ResNet50, aiming to extract both 2D local and 3D global features to classify COVID-19, CAP, and non-pneumonia.
Xu et al. \cite{xuh1n1} introduced a location-attention model to categorize COVID-19, Influenza-A viral pneumonia, and healthy cases. It took the relative distance-from-edge of segmented lesion candidates as extra weight in a fully connected layer to offer distance information.

Recently, Vision Transformer (ViT) has demonstrated its potentials by achieving competitive results on a variety of computer vision tasks. Relevant studies have also been conducted on the COVID-19 diagnosis. Gao et al. \cite{gao2021covid} used a ViT based on the attention models to classify COVID and non-COVID CT images. 
To integrate the advantages of convolution and transformer for COVID-19 detection, Park et al. \cite{park2021vision} presented a novel architecture that utilized CNN as a feature extractor for low-level Chest X-ray feature corpus, upon which Transformer was trained for downstream diagnosis tasks with the self-attention mechanism.

\subsection{Advanced network architecture}
In our approach, we adopt two representative deep learning architectures as the backbones, namely ResNeSt-50 and Uniformer-S. Here, we briefly introduce the closely related ResNet and Transformer architectures and their variants.

In the family of ResNets, ResNet \cite{he2016resnet} introduced a deep residual learning framework to address the network degradation problem.
ResNeXt \cite{xie2017resnext} established a simple architecture by adopting group convolution in the ResNet bottleneck block.
ResNeSt \cite{zhang2020resnest} presented a modular split-attention block within the individual network blocks to enable attention across feature-map groups.

Although CNN models have shown promising results, the limited receptive field makes it hard to capture global dependency. To solve this problem, Vision Transformer (ViT) \cite{dosovitskiy2020image} was applied to the sequences of image patches for an image classification task. Later on, Swin Transformer \cite{liu2021swin} proposed to use shifted windows between consecutive self-attention layers, which had the flexibility to model at various scales and had linear computational complexity with respect to the image size. 
Multi-scale Vision Transformer \cite{fan2021multiscale} connected the seminal idea of multi-scale feature hierarchies with transformer models for video and image recognition.
Pyramid Vision Transformer \cite{wang2021pyramid} used a progressive shrinking pyramid to reduce the computations of large feature maps, which overcame the difficulties of porting Transformer models to various dense prediction tasks and inherits the advantages of both CNN and Transformer. 
Unified transformer (UniFormer) \cite{li2022uniformer} sought to integrate the merits of convolution and self-attention in a concise transformer format, which can tackle both local redundancy and global dependency. To achieve the balance between accuracy and efficiency, we adopt Uniformer as the default backbone network. 

\section{Methodology} \label{section:method}

\begin{figure}[t]
\centering
\includegraphics[width=\textwidth]{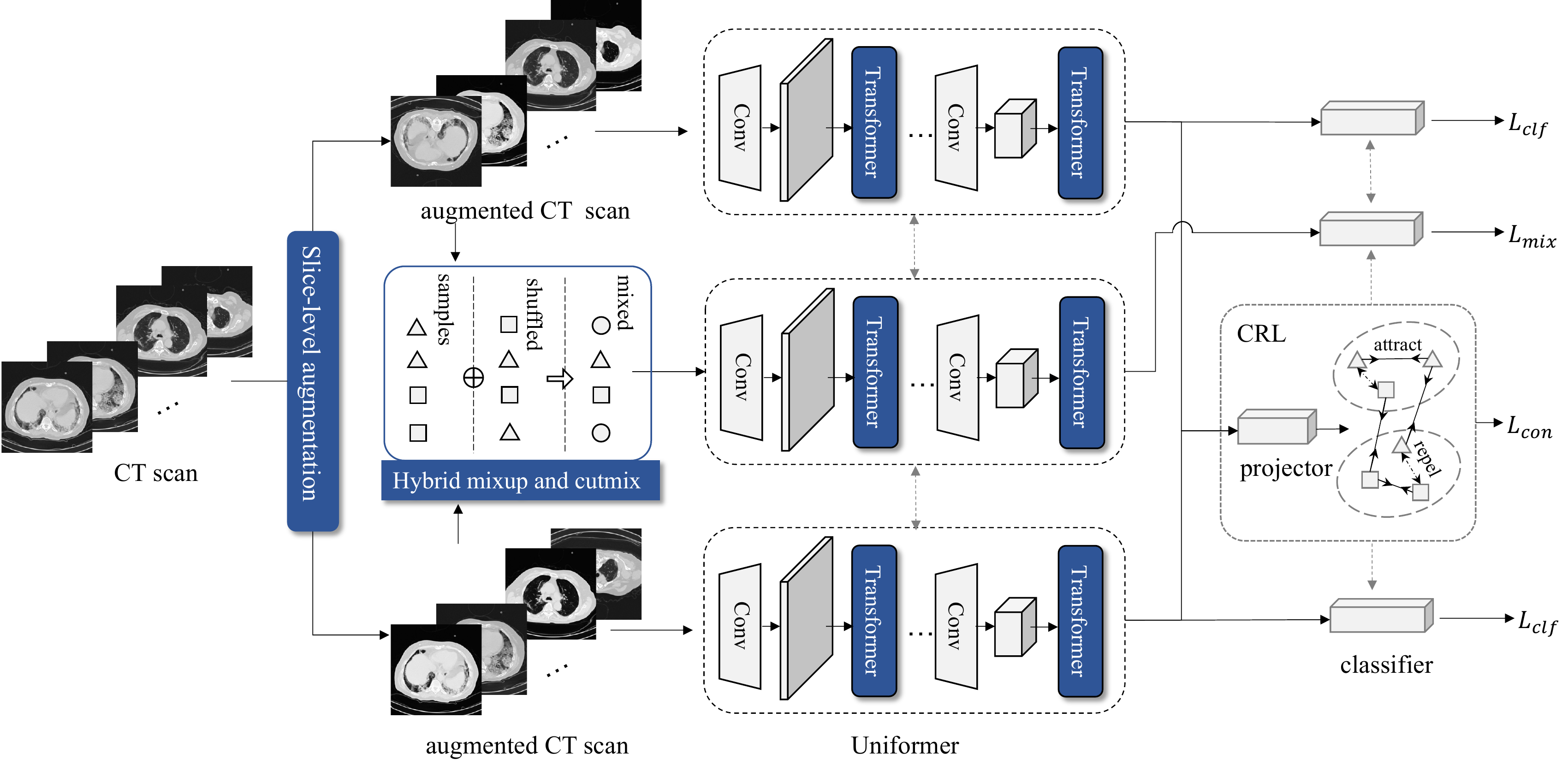}
\caption{Overview of our CMC\_v2 network for COVID-19 detection.}
\label{fig:cmc}
\end{figure}


The overall framework of our model is shown in Fig. \ref{fig:cmc}. In this section, we review the baseline method CMC\_v1 \cite{hou2021cmc} firstly and then introduce several simple and effective mechanisms to boost the detection performance. 

\subsection{Recap of CMC\_v1} 
CMC\_v1 employs the contrastive representation learning (CRL) as an auxiliary task to learn discriminative representations of COVID-19. CRL is comprised of the following components. 1) A stochastic data augmentation module $A(\cdot)$, which transforms an input CT $x_i$ into a randomly augmented sample $\tilde{x}_i$. Two augmented volumes are generated from each input CT scan. 2) A base encoder $E(\cdot)$, mapping the augmented CT sample $\tilde{x}_i$ to its feature representation $r_i=E(\tilde{x}_i)\in \mathbb{R}^{d_e}$. 3) A projection network $P(\cdot)$, used to map the representation vector $r_i$ to a relative low-dimension vector $z_i=P(r_i)\in \mathbb{R}^{d_p}$. 4) A classifier $C(\cdot)$, classifying the vector $r_i\in \mathbb{R}^{d_e}$ to the final prediction. 

\subsubsection{Contrastive representation learning.}
Given a minibatch of $N$ CT volumes and their labels $\{(x_i,y_i)\}$, we can generate a minibatch of $2N$ samples $\{(\tilde{x}_i,\tilde{y}_i)\}$ after data augmentations. 
Inspired by the supervised contrastive loss \cite{2020Supervisedcon}, we define the positives as any augmented CT samples from the same category, whereas the CT samples from different classes are considered as negative pairs. Let $i\in \{1,\dots,2N\}$ be the index of an arbitrary augmented sample, the contrastive loss function is defined as:
\begin{equation}
    \mathcal{L}_{con}^i=\frac{-1}{2N_{\tilde{y}_i}-1}\sum_{j=1}^{2N}\mathbbm{1}_{i\ne j}\cdot\mathbbm{1}_{\tilde{y}_i=\tilde{y}_j}\cdot\log\frac{\exp(z_i^T\cdot z_j/\tau)}{\sum_{k=1}^{2N}\mathbbm{1}_{i\ne k}\cdot\exp(z_i^T\cdot z_k/\tau)},
\label{infonce}
\end{equation}
where $\mathbbm{1}\in\{0,1\}$ is an indicator function, and $\tau >0$ denotes a scalar temperature hyper-parameter. $N_{\tilde{y}_i}$ is the total number of samples in a minibatch that have the same label $\tilde{y}_i$. 

\subsubsection{Mixup classification.}
CMC\_v1 adopts the mixup \cite{zhang2017mixup} strategy during training to further boost the generalization ability of the model. For each augmented CT sample $\tilde{x}_i$, the mixup sample and its label are generated as:
\begin{equation}
    \tilde{x}^{mix}_i = \lambda\tilde{x}_i + (1-\lambda)\tilde{x}_{p}, 
    ~\tilde{y}^{mix}_i = \lambda\tilde{y}_i + (1-\lambda)\tilde{y}_{p}, 
\end{equation}
where $p$ is randomly selected indice; $\lambda$ is the balancing coefficient. The mixup loss is defined as the cross-entropy loss of mixup samples:
\begin{equation}
    \mathcal{L}^i_{mix} = \mathrm{CrossEntropy}(\tilde{x}^{mix}_i,\tilde{y}^{mix}_i).
\end{equation}

Different from the original design \cite{zhang2017mixup} where they replaced the classification loss with the mixup loss, we merge the mixup loss with the standard cross-entropy classification loss $\mathcal{L}^i_{clf}= \mathrm{CrossEntropy}(\tilde{x}_i,\tilde{y}_i)$ to enhance the classification ability on both mixup samples and raw samples. 

The total loss is defined as the combination of the contrastive loss, mixup loss, and classification loss:
\begin{equation}
     \mathcal{L}=\frac{1}{2N}\sum_{i=1}^{2N}(\mathcal{L}^i_{con}+\mathcal{L}^i_{mix}+\mathcal{L}^i_{clf}).
\end{equation}

\subsection{Improving COVID-19 detection with CMC\_v2}
To boost the COVID-19 detection performance, we incorporate natural video priors into CMC\_v1 by adapting an efficient pre-trained video backbone to our task, and develop a hybrid data augmentation strategy to increase data efficiency.

\subsubsection{Transfer learning with a stronger backbone and pre-training.}
In CMC\_v1, a 3D ResNeSt-50 model \cite{zhang2020resnest} is employed as the backbone network for feature extraction. Although 3D convnets capture local volume semantics efficiently, they are incapable of modeling long-range dependencies between spatial/temporal features explicitly. For simplicity, we refer `temporal' to the relationship among different CT slices in this paper. Recent works on Vision Transformer \cite{dosovitskiy2020image} managed to encode long-range information using self-attention. However, global self-attention is computationally inefficient and transformer models only demonstrate superior results when huge data is available. Compared with natural image datasets, COVID-19 image datasets have a smaller scale and the model is prone to overfitting. To alleviate this issue, we adopt a video transformer named Uniformer \cite{li2022uniformer}, a novel hybrid CNN-transformer model which integrates the advantages of convolution and self-attention in spatial-temporal feature learning while achieving the balance between accuracy and efficiency. In particular, Uniformer replaces the naive transformer block with a Uniformer block, which is comprised of a Dynamic Position Embedding (DPE) layer, a Multi-Head Relation Aggregator (MHRA) layer, and a Feed-Forward Network (FFN). 

Furthermore, we experimentally find that training the model from scratch leads to poor results. In transfer learning, it is a common practice to initialize the model on downstream tasks with weights pre-trained on a large-scale ImageNet dataset. To initialize the 3D model, CMC\_v1 inflated the ImageNet pre-trained 2D weights to the 3D model. This is achieved by either copying the 2D weights to the center of the 3D weights or repeating the 2D weights along the third dimension. However, these inflated 3D weights may not excel at modeling the temporal relationship between different slices. To address this issue, we directly initialize the model with 3D weights pre-trained on video action recognition datasets, i.e. k400 \cite{k400}. We empirically prove that k400 pre-training yields better results than inflated weight initialization in this task. 

\subsubsection{Hybrid mixup and cutmix strategy.}
In CMC\_v1, the mixup strategy is introduced to generate diverse CT samples. These pseudo samples are beneficial for improving the model's generalization ability. Similar to a mixup, cutmix \cite{yun2019cutmix} replaces a local region in the target image with the corresponding local region sampled in the source image. To combine the merits of both, we develop a hybrid mixup and cutmix strategy. In each iteration, we select one strategy with equal probability. This hybrid strategy works well on the traditional Data Parallel (DP) mechanism \cite{paszke2019pytorch} in multi-GPU training. However, it's challenging to scale to the modern Distributed Data Parallel (DDP) mechanism. The original batch size on each GPU is set to 1 in our case because the effective batch size is 4 after two-view augmentation and hybrid mixup and cutmix strategy, reaching the memory limit on each GPU (The shape of the mini-batch tensors on each GPU is 4$\times$T$\times$3$\times$H$\times$W). As the DDP mechanism starts an individual process on each GPU, the hybrid strategy is directly employed on each GPU individually. Performing the hybrid mixup and cutmix strategy on the augmented views of the same image does not align with the original effect. To make it work, we gather all the samples from the GPUs, conduct the hybrid mixup and cutmix over all the samples, and dispatch the generated samples back to each GPU. It guarantees that the hybrid strategy is performed across different CT scans in the current mini-batch. The hybrid mixup and cutmix strategy boost the model's generalization ability.

\subsubsection{Slice-level augmentation.}
The data augmentation strategies used in CMC\_v1 are 3D rescaling, 3D rotation, and color jittering on all the slices. To further increase the data diversity, we follow the common practice in video data processing and perform different 2D augmentations on each slice, termed as SliceAug. SliceAug achieves slightly better performance than 3D augmentation while having a comparable pre-processing time. 

\subsubsection{Small resolution training.}
Prior works \cite{dosovitskiy2020image,touvron2021training} have demonstrated the effectiveness of using small image resolution during training and large resolution during validation/testing. This mechanism bridges the gap between the image size mismatch caused by the random resized cropping during training and center cropping during testing \cite{touvron2019fixing}. Besides, the small resolution makes training more efficient. In the experiments, we use the resolution of 192$\times$192 and 224$\times$224 for training and testing, respectively. 

\section{Dataset} \label{section:dataset}

We evaluate our proposed approach on the COV19-CT-Database (COV19-CT-DB) \cite{kollias2022ai}. The COV19-CT-DB contains chest CT scans marking the existence of COVID-19. It consists of about 1,650 COVID and 6,100 non-COVID chest CT scan series from over 1,150 patients and 2,600 subjects. In total, 724,273 slices correspond to the CT scans of the COVID category and 1,775,727 slices correspond to the non-COVID category.
Data collection was conducted in the period from September 1, 2020 to November 30, 2021. 
Annotation of each CT scan was obtained by 4 experienced medical experts and showed a high degree of agreement (around 98\%). Each 3D CT scan includes a different number of slices, ranging from 50 to 700. This variation in the number of slices is due to the context of CT scanning. 
The database is split into training, validation, and testing sets. The training set contains 1,992 3D CT scans (1,110 non-COVID cases and 882 COVID cases). The validation set consists of 504 3D CT scans (289 non-COVID cases and 215 COVID cases). The testing set includes 5,281 scans and the labels are not available during the challenge.     


\section{Experiments} \label{section:experiments}

\subsection{Implementation details}

All CT volumes are resized from $(T,512,512)$ to $(128,224,224)$, where $T$ denotes the number of slices. For training, data augmentations include random resized cropping on the transverse plane, random cropping on the vertical section to $64$, rotation, and color jittering. 
We employ the 3D ResNeSt-50 and Uniformer-S as the backbones in our experiments. The value of parameter $d_e$ is 2,048/512 for ResNeSt-50/Uniformer-S, and $d_p$ is set to 128. 
All networks are optimized using the Adam algorithm with a weight decay of 1e-5. The initial learning rate is set to 1e-4 and then divided by 10 at $30\%$ and $80\%$ of the total number of training epochs. The networks are trained for 100 epochs. Our methods are implemented in PyTorch and run on eight NVIDIA Tesla A100 GPUs.

\subsection{Evaluation metrics}
To evaluate the performance of the proposed method, we adopt the same official protocol of 2nd COVID-19 Competition as the evaluation metric. We report F1 Scores for non-COVID and COVID categories as well as the Macro F1 Score for overall comparison. The Macro F1 Score is defined as the unweighted average of the class-wise/label-wise F1 Scores. We also present ROC curves and Area Under Curve (AUC) for each category.

\subsection{Ablation studies on COVID-19 detection challenge}

\setlength{\tabcolsep}{4pt}
\begin{table}[t]
\begin{center}
\caption{The results on the validation set of COVID-19 detection challenge.} 
\label{table:covid detection}
\resizebox{\linewidth}{!}{
\begin{tabular}{llcccccc}
\hline\noalign{\smallskip}
\multirow{2}{*}{ID} & \multirow{2}{*}{Method} & \multirow{2}{*}{Param} & \multirow{2}{*}{FLOPs} & \multirow{2}{*}{Pre-train}  & \multirow{2}{*}{Macro F1} & \multicolumn{2}{c}{F1}\\
\cline{7-8}
\noalign{\smallskip}
&&&&&& Non-COVID  & COVID\\
\noalign{\smallskip}
\hline
\noalign{\smallskip}
1& ResNet50-GRU \cite{kollias2022ai} & - & - & - & 77.00 & - & -\\
2& ResNeSt-50 & 52.8M & 371.9G & ImageNet & 89.89 & 91.27 & 88.52 \\
3& Uniformer-S & 21.2M & 230.1G & ImageNet & 90.98 & 92.08 & 89.88\\
4& CMC\_v1 (R) & 57.3M & 371.9G & ImageNet &  91.98 & 93.14 & 90.82\\
5& CMC\_v1 (U) & 21.5M & 230.1G & ImageNet &  92.26 & 93.11 &   91.42\\
6& CMC\_v1 (U) & 21.5M & 230.1G & k400\_16$\times$4 & 92.48 & 93.28 & 91.67\\
7& CMC\_v1 (U) & 21.5M & 230.1G & k400\_16$\times$8 & 92.70 & 93.41 & 91.99\\
\noalign{\smallskip}
\hline
\noalign{\smallskip}
8& CMC\_v2 (U, SliceAug) & 21.5M & 230.1G & k400\_16$\times$8 & 93.07 & 93.94 & 92.20\\
9& CMC\_v2 (U, Hybrid) & 21.5M & 230.1G & k400\_16$\times$8 & 93.29 & 94.12 & 92.45\\
10& CMC\_v2 (U, Hybrid+SmallRes) & 21.5M & 169.1G & k400\_16$\times$8 & 93.30 & 94.07 & 92.52 \\
\hline
\end{tabular}
}
\end{center}
\end{table}
\setlength{\tabcolsep}{1.4pt}

We conduct ablation studies on the validation set of COVID-19 detection challenge to show the impact of each component of our proposed methods. We first analyze the effects of different backbones, and then we discuss the effectiveness of the CMC\_v1 framework and the choice of various pre-training methods. Finally, we investigate the impact of the new components in our CMC\_v2, i.e. slice-level augmentation (SliceAug), hybrid mixup and cutmix strategy (Hybrid), and small resolution training (SmallRes).

\subsubsection{Backbone network.}
To analyze the effects of architectures, we compare different backbone models, and the results are shown in the first three rows of Table \ref{table:covid detection}. The reported result of the baseline approach `ResNet50-GRU' \cite{kollias2022ai} is 77.00\% Macro F1 Score. This model is based on CNN-RNN architecture \cite{kollias2020deep,kollias2020transparent,kollias2018deep}, where the CNN part performs local analysis on each 2D slice, and the RNN part combines the CNN features of the whole 3D CT scan. 
Compared to the baseline, our 3D ResNeSt-50 and Uniformer-S backbones achieve more than 12\% improvements on the Macro F1 Scores. Specifically, the Uniformer-S achieves better performance on all the metrics, surpassing ResNeSt-50 by 1.09\% Macro F1 Score, 0.81\% and 1.36\% F1 Scores for non-COVID and COVID classes. Besides, the Uniformer-S greatly reduces the network parameters and computational costs. The results demonstrate the long-range dependencies modeling ability of Uniformer-S, which is important to capture the relationships between different CT slices. 

\subsubsection{Analysis of CMC\_v1.}
We evaluate the effectiveness of the previous CMC\_v1 network. The 4th and 5th rows in Table \ref{table:covid detection} show the results of CMC\_v1 (R) and CMC\_v1 (U), where the R and U denote ResNeSt-50 and Uniformer-S backbones, respectively. CMC\_v1 on both backbones can achieve significant performance improvements. In particular, CMC\_v1 (U) obtains 92.26\% on Macro F1 Score, 93.11\% and 91.42\% on F1 Scores for non-COVID and COVID categories. The results demonstrate the generality of the CMC\_v1, which can consistently improve the COVID-19 detection performance with different backbones. 

\subsubsection{Pre-training schemes.}

We compare three pre-training methods, namely ImageNet, k400\_16$\times$4, and k400\_16$\times$8. ImageNet pre-training inflates the 2D pre-trained weights to our 3D models. K400 pre-training denotes 3D weights pre-trained on the video action recognition dataset k400, where 16$\times$4 and 16$\times$8 indicate the sampling 16 frames with frame stride 4 and 8, respectively. It can be seen from the 5th to 7th rows in Table \ref{table:covid detection}, CMC\_v1 (U) with k400\_16$\times$8 pre-training weights outperforms the other two methods on all metrics. Based on the above results, we choose the Uniformer-S with k400\_16$\times$8 pre-training weights as the default backbone for our proposed CMC\_v2. 

\begin{figure}[t]
\centering
\includegraphics[width=\textwidth]{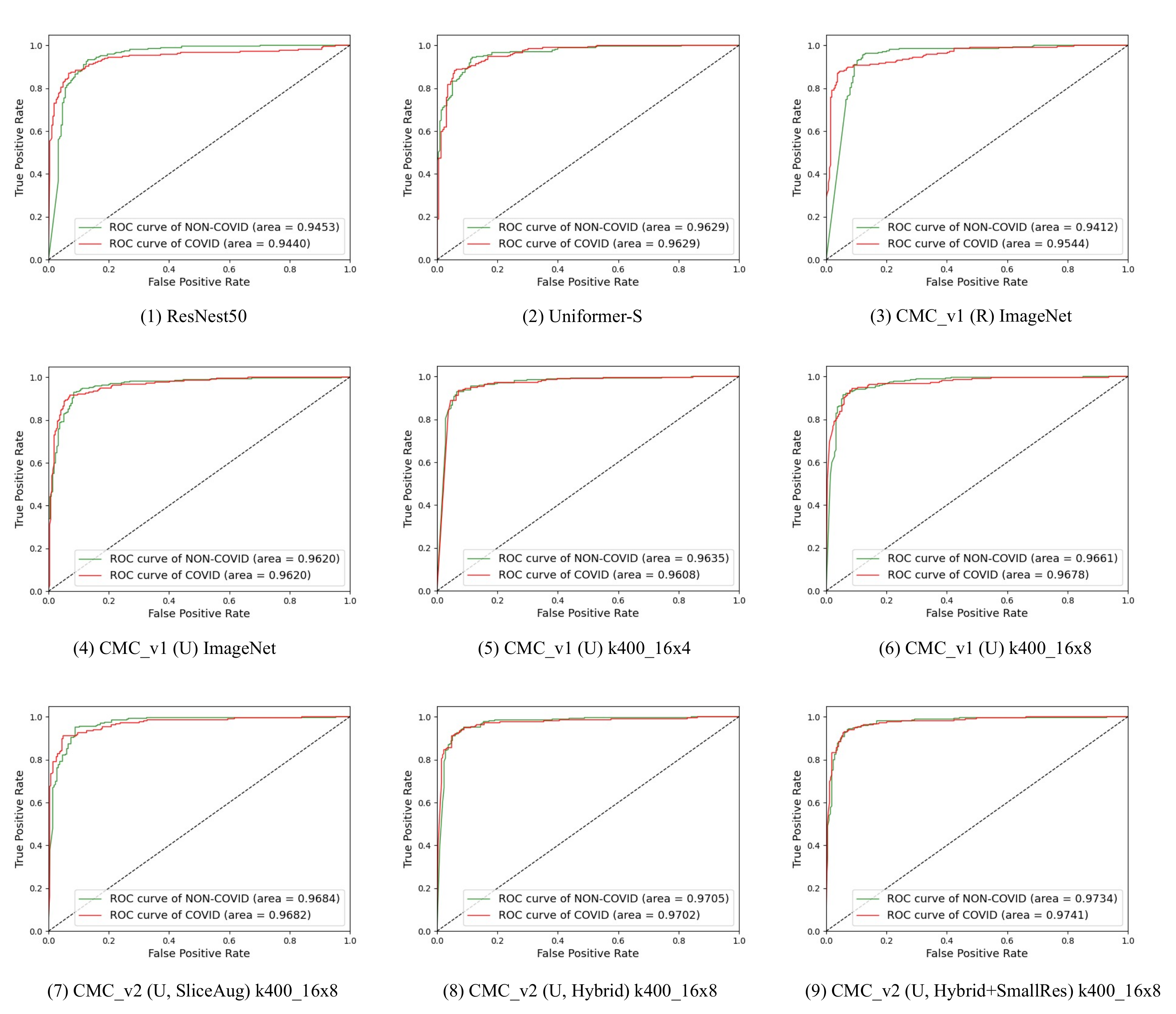}
\caption{The ROC curves and AUC scores of different networks.}
\label{fig:roc}
\end{figure}

\subsubsection{Analysis of CMC\_v2.}

In this part, we investigate the impact of our newly proposed components in CMC\_v2, including slice-level augmentation (SliceAug), hybrid mixup and cutmix strategy (Hybrid), and small resolution training (SmallRes). 
The experimental results in the 8th row of Table \ref{table:covid detection} indicate that CMC\_v2 (U, SliceAug) can improve the performance on all metrics compared with the CMC\_v1 (U). The slice-level augmentation can further increase the data diversity and benefit COVID-19 detection performance. 
As for the hybrid mixup and cutmix strategy, the CMC\_v2 (U, Hybrid) achieves further improvement by 0.59\% Macro F1 Score, 0.71\% COVID F1 Score, and 0.46\% non-COVID F1 Score, compared with the CMC\_v1 (U) that only employs the single mixup strategy. Our hybrid mixup and cutmix strategy generates diversified data for improving the model's generalization ability in COVID-19 detection.
When we adopt the small resolution training mechanism, the CMC\_v2 (U, Hybrid+SmallRes) achieves the best performance with minimal computational costs among all models. It obtains 93.30\% on Macro F1 Score, 94.07\% on non-COVID F1 Score, and 92.52\% on COVID F1 Score. In particular, this model shows the superior recognition ability for the COVID-19 category among all other approaches. 

In addition, we present the ROC curves and AUC of our models in Fig. \ref{fig:roc}. The AUC results of all the models reach more than 0.94 for both non-COVID and COVID classes. Especially, the full version of CMC\_v2 (U, Hybrid+SmallRes) obtains the highest AUC Scores (0.9734 and 0.9741 for non-COVID and COVID, respectively) among all settings.

\subsection{Results on COVID-19 detection challenge leaderboard}

Table \ref{table:test} shows the results of our method and other participants on the testing set of 2nd COVID-19 detection challenge. Our method ensembles all the CMC\_v2, including CMC\_v2 (U, SliceAug), CMC\_v2 (U, Hybrid), and CMC\_v2 (U, Hybrid+SmallRes) following the strategy in \cite{hou2021cmc}. The final prediction of each CT scan is obtained by averaging the predictions from individual models. We also adopt a test time augmentation (TTA) operation to boost the generalization ability of our models on the testing set. It can be seen from Table \ref{table:test} that our proposed method
ranks first in the challenge with 89.11\% Macro F1 Score. Compared to other methods, our model achieves significant improvement on the F1 Score for the COVID category (80.92\%), indicating the ability to distinguish COVID cases from non-pneumonia and other types of pneumonia correctly.

\setlength{\tabcolsep}{4pt}
\begin{table}[t]
\begin{center}
\caption{The leaderboard on the 2nd COVID-19 detection challenge.}
\label{table:test}
\begin{tabular}{p{30pt}p{80pt}p{60pt}<{\centering}p{60pt}<{\centering}p{60pt}<{\centering}}
\hline\noalign{\smallskip}
\multirow{2}{*}{Rank} & \multirow{2}{*}{Teams} & \multirow{2}{*}{Macro F1} & \multicolumn{2}{c}{F1}\\
\cline{4-5}
\noalign{\smallskip}
&&& Non-COVID & COVID\\
\noalign{\smallskip}
\hline
\noalign{\smallskip}
1& FDVTS (Ours) & 89.11 & 97.31 & 80.92\\
1& ACVLab & 89.11 & 97.45 & 80.78\\
3& MDAP & 87.87 & 96.95 & 78.80 \\
4& Code 1055 & 86.18  & 96.37  & 76.00 \\
5& CNR-IEMN & 84.37 &  95.98 & 72.76\\
6& Dslab & 83.78 & 96.22 & 71.33 \\
7& Jovision-Deepcam & 80.82 &  94.56 & 67.07 \\
8& ICL & 79.55 & 93.77 & 65.34 \\
9& etro & 78.72 & 93.48 & 63.95 \\
10& ResNet50-GRU \cite{kollias2022ai} & 69.00 & 83.62 & 54.38 \\
\hline
\end{tabular}
\end{center}
\end{table}
\setlength{\tabcolsep}{1.4pt}

\subsection{Visualization results}

\begin{figure}[t]
\centering
\includegraphics[width=\textwidth]{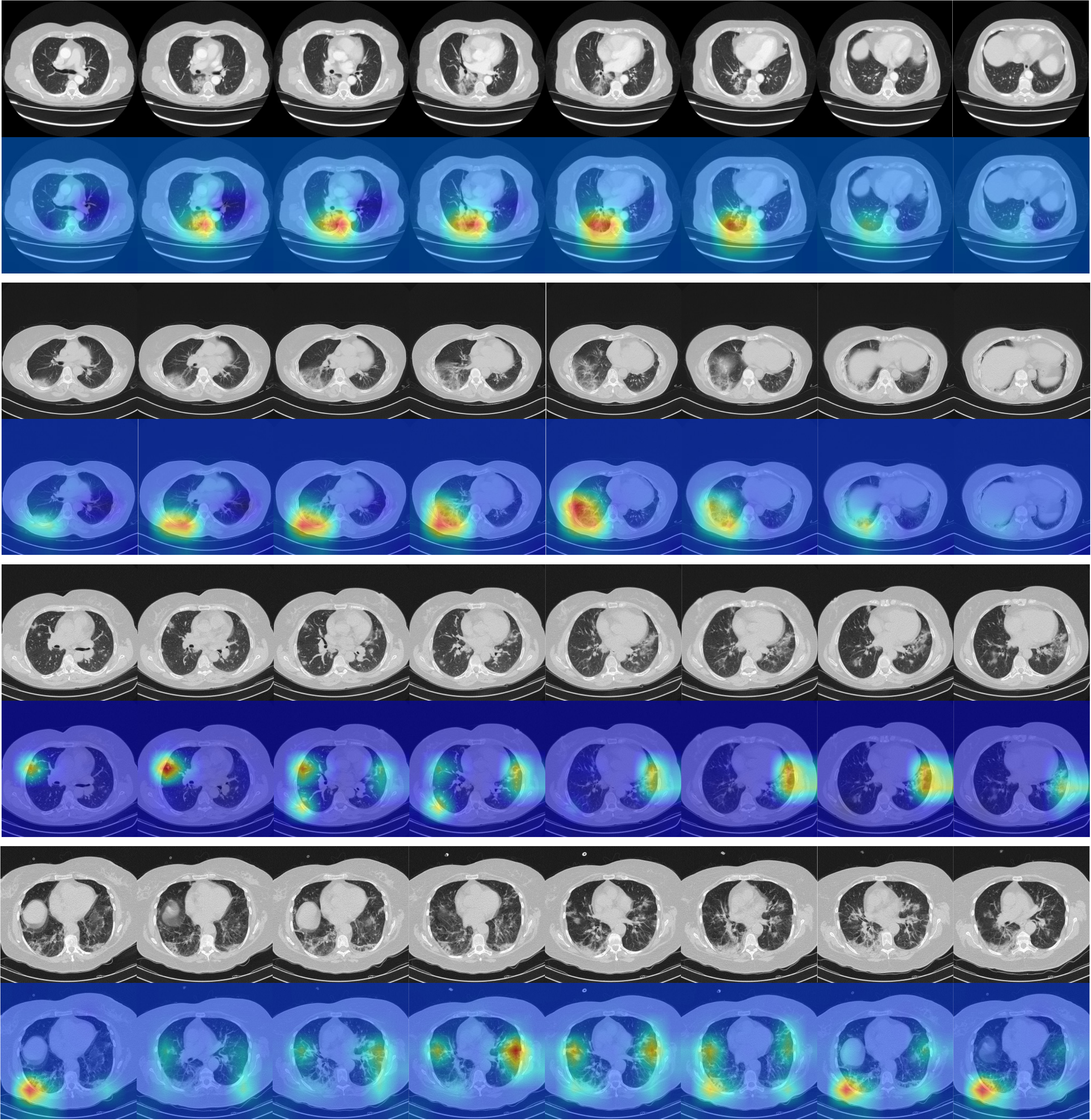}
\caption{The visualization results on the COVID-19 CT scans.}
\label{fig:cam}
\end{figure}

To verify the interpretability of our model, we visualize the results using Class Activation Mapping (CAM) \cite{zhou2016learning}. As illustrated in Fig. \ref{fig:cam}, we select four COVID-19 CT scans from the validation set of COV19-CT-DB dataset. In each group, the upper row shows the series of CT slices, and the lower row presents the corresponding CAM results. In the first group, it can be seen that the attention maps focus on the local infection regions accurately. 
In the second group, the wide range of infection regions can also be covered. In the third and fourth groups, the infections in bilateral lungs can also be located precisely. 
These attention maps provide convincing interpretability for the COVID-19 detection results, which is helpful for real-world clinical diagnosis.

\section{Conclusions} \label{section:conclusion}
In this paper, we propose a novel and practical solution winning COVID-19 detection at the 2nd COVID-19 Competition. Based on the CMC\_v1 network, we further develop the CMC\_v2 network with substantial improvements, including the CNN-transformer video backbone, hybrid mixup and cutmix strategy, slice-level augmentation, and small resolution training mechanism. The experimental results demonstrate that the new components boost the COVID-19 detection performance and the generalization ability of the model. 
On the testing set, our method ranked 1st in the 2nd COVID-19 Competition with 89.11\% Macro F1 Score among 14 participating teams.

\section*{Acknowledgement}
This work was supported by the Scientific \& Technological Innovation 2030 - ``New Generation AI'' Key Project (No. 2021ZD0114001; No. 2021ZD0114000), and the Science and Technology Commission of Shanghai Municipality (No. 21511104502; No. 21511100500; No. 20DZ1100205). Yuejie Zhang, Xiaobo Zhang, and Rui Feng are corresponding authors.

%
%
\bibliographystyle{splncs04}
\bibliography{egbib}
\end{document}